\documentclass{article}
   \usepackage{graphicx}

\makeatletter
\@addtoreset{figure}{section}
\def\thefigure{\thesection.\@arabic\c@figure}
\def\fps@figure{h, t}
\@addtoreset{table}{section}
\def\thetable{\thesection.\@arabic\c@table}
\def\fps@table{h, t}
\@addtoreset{equation}{section}

\makeatother

\usepackage{pslatex}
\author{Kurt Ehlers\thanks{
           Corresponding author.  Address: Mathematics Department ,
	Truckee Meadows Community College, 7000 Dandini Blvd, Reno, NV 89512 (kehlers@tmcc.edu)} \\
	Truckee Meadows Community College \\
	7000 Dandini Blvd, Reno, NV 89512  \\
            \and Jair Koiller \\
	Mathematics Department,
	Funda\c c\~ao Getulio Vargas\\ Praia de Botafogo 190,
Rio de Janeiro, RJ, 22250-040,
 Brazil  }

 \title{ Synechococcus as a ``singing"  bacterium: biology inspired by micro-engineered acoustic streaming devices}

\begin{document}

\newcommand{\R}{\hbox{\bf R}}


\newtheorem{proposition}{Proposition}
\newtheorem{theorem}{Theorem}
\newtheorem{lemma}{Lemma}
\newtheorem{definition}{Definition}
\newtheorem{notation}{Notation}
\newtheorem{remark}{Remark}
\newtheorem{assertion}{Assertion}
\newtheorem{conjecture}{Conjecture}
\newtheorem{corollary}{Corollary}
\newtheorem{exercise}{Exercise}
\newtheorem{problem}{Problem}
\newtheorem{criterion}{Criterion}

\addtolength{\topmargin}{-.775in}
\addtolength{\textheight}{1.5in}
\newcommand{\vU}{{\bf U}}
\newcommand{\vX}{\vec{X}}
\newcommand{\vx}{\vec{x}}
\newcommand{\vV}{\vec{V}}
\newcommand{\vv}{\vec{v}}
\newcommand{\vn}{\vec{n}}
\newcommand{\vN}{\vec{N}}
\newcommand{\vu}{{\bf u}}
\newcommand{\vT}{\vec{T}}
\newcommand{\vF}{\vec{F}}
\newcommand{\vW}{\vec{W}}
\newcommand{\vc}{\vec{c}}
\newcommand{\vr}{\vec{ r}}
\newcommand{\vd}{\vec{ d}}
\newcommand{\va}{\vec{ a}}
\newcommand{\cd}{{\cal D}}
\newcommand{\ta}{\tilde{a}}
\newcommand{\tb}{\tilde{b}}
\newcommand{\tc}{\tilde{c}}
\newcommand{\T}{{\bf T}}
\newcommand{\ho}{\hat{\Omega}}
\newcommand{\hT}{\hat{T}}

\maketitle

\pagestyle{headings}

\noindent \footnotesize{ {\bf Abstract.}  Certain  cyanobacteria, such as open ocean strains of Synechococcus,
are able to swim at speeds up to 25 diameters per second, without flagella or visible changes in shape.
The means by which Synechococcus generates thrust for self-propulsion is unknown.
The only mechanism that has not been ruled out employs tangential waves of surface deformations.
In \cite{Ehlersetal} the average swimming velocity for this mechanism was estimated using the methods inaugurated by Taylor and Lighthill in the 1950's and revisited in differential geometric language by Shapere and Wilczek in 1989.  The procedure consists of solving  quasi-statically the  Stokes equations with no slip boundary conditions.  (These are given by the instantaneous velocity field defined by the current deformation of a localized shape. The physical condition of no net force and torque yields a rigid body counterflow, whose average on a stroke cycle gives  the average swimming velocity.)  In this article we propose making a break with the no slip boundary condition paradigm.   In fact,  we  are proposing here an entirely different  physical principle self propulsion based on  {\it acoustic streaming}.  Micro-pumps in silicon chips, based on AS,  have been constructed by engineers since the 1990's,  but to the best of our knowledge acoustic streaming as a means of  microorganisms locomotion  has not been proposed before. Our hypothesis is supported by two recent remarkable discoveries: (1)  In  \cite{Samuel},  deep-freeze electron microscopy of the motile strain WH8113  revealed a crystalline outer layer (CS) covered with a forest of  "spicules" (Sp) extending from the inner membrane through the CS, projecting 150 nm into the surrounding fluid.  (2)  In \cite{Pelling}, atomic force microscopy (AFM)  was used to find that the cell wall of yeast cells periodically oscillates on  nano-scale amplitudes at frequencies of 0.8 to 1.6 kHz,  and that the oscillations are generated metabolically. We propose that the spicules, in contact with the cell's power systems, could perform high frequency motions generating acoustic streaming (AS) in the surrounding fluid. We compare two models for self-propulsion employing acoustic streaming: the quartz wind effect (QW) and boundary induced streaming generated by surface acoustic waves (SAW). Based on an estimate of the power required, the former would require an enhancement mechanism similar to a laser to be viable. In striking contrast, we find that the efficiency of the SAW mechanism compares favorably with known strategies for bacterial self-propulsion. The required amplitude is below the resolution limit of light microscopy and the required frequency is biologically attainable. Moreover SAW produce an "atmosphere" (the Stokes layer) surrounding the cell, within which the fluid motion is essentially chaotic and thus acoustic streaming may turn out to be biologically advantageous, enhancing nutrient uptake and chemical reactions.  Some possible experiments are outlined.
}
\\ 

\\


{\small

\section{Introduction}
 The aim of this note is to explore the possibility that one of the current strategies used in micro-engineering, {\it acoustic streaming},  that
could  have been emulated by Nature around 4 billion years ago.  By acoustic streaming we refer to the mean flow in a fluid associated with the attenuation of an acoustic wave.

Motile strains of Synechococcus were discovered in the Atlantic ocean in 1985  \cite{waterbury} and are featured in recent reviews \cite{Jarrell}, \cite{Bra1} . Swimming at speeds of 10-25 diameters per second, their locomotion is unusual in that it does not involve flagella  or other structures typically associated with bacterial motility and the means by which they generate thrust for self-propulsion remains a mystery \cite{willey}. Both sodium and calcium are required for motility \cite{pittao}.

Our goal is to explore whether it is theoretically possible for Synechococcus to generate thrust for propulsion using acoustic streaming. With the recent developments in technologies such as Atomic Force Microscopy (AFM) \cite{Pelling} and Total Internal Reflection Microscopy \cite{Guasto} we believe the time is ripe to solve the mystery of Synechococcus motility.

Our main results are to show that while the simplest mechanism involving acoustic streaming, the so called quartz wind, is too inefficient to generate the observed speeds a boundary induced streaming mechanism involving a traveling surface acoustic wave has efficiency that compares favorably with other strategies observed in nature. We emphasize that the acoustic streaming models are fundamentally different from the squirming mechanism. The latter involves purely linear fluid mechanics at its roots, while the former is ab-initio a nonlinear effect.

\subsection{Acoustic streaming powered MEMS devices and living counterparts}
In a recent review on micro-electromechanical (MEMS) devices, Squires and Quake   \cite{SQ} classify  the physical processes for micro-engineered fluid streaming into three main types: A. {\it Electrokinetic } (electroosmosis/electrophoresis), B. {\it Steady streaming}
 (in particular, acoustic streaming resulting from the gradient of Reynolds stresses),  and  C. {\it Fluid structure interactions}.  The latter involves  membrane deformations, e.g.., the use of soft polymeric tunable  materials.

 In this paper we focus on type B. The application of acoustic streaming, via surface acoustic waves (explained below) to micro-fluidic pumping and mixing devices was pioneered by Moroney, White and Howe in 1991 \cite{Moroney}.  A new generation of MEMS pumps and valves based on surface acoustic waves(SAW) producing net streaming in micro-platforms and channels are now available  \cite{Cecchini}, \cite{Hashimoto}, \cite{Lutz}, \cite{Marmottant1}, \cite{Marmottant}, \cite{Rathgeber}, \cite{Renaudin1},
 \cite{Renaudin}, \cite{rife}, \cite{Sritharan}, \cite{suria}, \cite{Tan}, \cite{Tan1}, \cite{Wixforth}.

``Test pilots''  to compare these possible strategies in Nature are  open-ocean isolates of Synechococcus  capable of swimming.  (Strains living in nutrient rich coastal waters are nonmotile.)  Type A mechanisms such as electrophoresis  have been ruled out both experimentally and on physical grounds \cite{pitta}. In this model the cell carries a fixed charge that is shielded by counter-ions in the surrounding fluid. The organism pumps charged ions from one end of the cell and absorbs them at the other creating an electric field in the surrounding fluid. This field creates a flow in the fluid containing the counter-ions propelling the cell.  This leaves the cell's outer membrane as the source for generating thrust necessary for propulsion.  Compression-expansion tangential surface waves along the membrane (squirming), a type C mechanism,  was proposed in the mid 90's \cite{Ehlersetal},
\cite{Stone}.  Motivated by the discovery that some cells are able to generate oscillatory motions on their outer membrane at acoustic frequencies,  we explore the possibility of
type B mechanisms involving acoustic streaming. (Actually, acoustic streaming was hinted in  \cite{rife} without further discussion.)

 While photosynthetic, motile strains of Synechococcus do not show a phototactic or photophobic response to light but do show a chemotactic response to certain nitrogenous compounds
\cite{waterbury1}.
Motility is thought to allow open-ocean Synechococcus to take advantage of micro-environments on the scale of millimeters or centimeters. The effect that we propose here may also enhance nutrient uptake
and chemical reactions. See the discussion section for further elaboration on this point.

Regarding the locomotion machinery, a new clue came from a recent electron microscopy study  \cite{Samuel}. Motile strains of Synechococcus have a crystalline outer shell (CS) whose component parts are arranged in a rhomboid lattice penetrated by a profusion of tiny spicules (Sp)  emerging from the outer membrane up to 150$nm$ into the surrounding fluid.  The spicules penetrate the inner membrane where electro-chemical energy is available to drive the propulsion system.  {\it The shell is lacking in non-motile strains} \cite{McCarren}.
This discovery, together with new evidence that cells are able to generate high frequency vibrations on their outer membrane prompted us to suggest that a type B mechanism, acoustic streaming,   involving  Sp-CS interactions, could explain the locomotion  of Synechococcus. We hypothesize that the spicules could actuate like piezoelectric drivers. Free to move in the fluid, the cell will swim rather than pump fluid.

We believe that many other cells may also use acoustic streaming for locomotion or to enhance fluid mixing near the membrane. Acoustic streaming could be involved in the gliding motions of other cyanobacteria, or in the self-propulsion of eukaryotes possessing a silica shell such as diatoms with a raphe micro-channel. High frequency oscillations of the outer membrane of cells are not unprecedented in Nature and, in fact, may be quite widespread.

 Using atomic force microscopy (AFM), the outer membranes of Yeast cells have been  observed to oscillate at between 0.8-1.6 KHz with typical amplitudes of $\sim$3nm \cite{Pelling}.   The oscillations were shown to be metabolically driven by molecular motors. It was demonstrated that they are not caused by brownian motion nor by artifacts of the AFM apparatus. The magnitude of the forces at the cell wall were measured to be $\sim$10nN suggesting that they are generated by many protein motors working cooperatively.  A theoretical model for how cells can generate high frequency oscillations using coupled molecular motors has been developed by Frank J\"{u}licher \cite{jul}.  The model predicts that molecular motors working in unison can achieve frequencies of 10KHz and beyond.

 In an effort to explain the ear's remarkable ability to sense sound over a range of six orders of magnitude in frequency and twelve orders of magnitude in intensity starting at sounds whose energy per cycle is less than that of thermal noise ($kT$), researchers speculate that structures within the inner ear spontaneously oscillate at acoustic frequencies to amplify weak signals \cite{jul}, see also the review article \cite{duke}. There is  experimental evidence to support this theory, see \cite{martin}.
Further, the rotary motors of e-coli, which are large membrane embedded structures, have been observed to rotate at 300Hz under no load conditions \cite{bergprivate}.

\subsection { Dynamic streaming vs. kinematic streaming.}  The basic principle underlying  swimming at the microscopic scale is that a neutrally buoyant, free swimming organism does not exert net forces or torques on the surrounding fluid \cite{Ludwig}. This condition is to be met at each instant. In the 1950's and 60's envelope deformation (squirming) models were developed by Taylor, Lighthill, and Blake and applied to ciliary propulsion.  For certain ciliates, those whose cilia tips remain close during the "power stroke," this model does a good job at predicting swimming speeds; Opalina provides an example. For ciliates such as Paramecia, the cilia tips are do not remain close during the power stroke and the squirming model under-predicts the swimming velocity by an order of magnitude. Paramecia are best thought of as rowers. The squirming model is similar to the models based on acoustic streaming presented here in that small amplitude cyclical motions along the outer membrane are rectified into mean streaming flow leading to large scale motion. The difference is in the underlying Physics leading to the rectification of the oscillating wave into linear motion.

For the squirming mechanism, cyclic but non-reciprocal shape changes on a virtual surface  generate an effective net motion through the fluid. Associated to an infinitesimal boundary motion, represented by a vector field there is a  there is a corresponding infinitesimal translation and rotation. Inertia plays absolutely no role in this theory and the motion associated with a swimming stroke is independent of the time taken in its execution. If the swimmer retraces the stroke in reverse it returns to its initial position; this is known as the {\it Scallop Theorem}.   The mathematical formulation involves a ``gauge theoretical framework'' so that cyclic motions on a shape space produce  holonomy  (a Euclidean motion after a cycle) in the space of located shapes (\cite{shapere}, see also \cite{Koiller1}).

  In the acoustic streaming models, mean flow is generated by  a  force that results from the attenuation of acoustic energy in the fluid.     Acoustic streaming is one instance where the nonlinear inertia terms in the Navier-Stokes equations play an important role even in low Reynolds number fluid mechanics. Streaming here is inherently dynamic.
   We follow Lighthill's review \cite{Lighthill} where he describes the theory in both the high and low Reynolds number regimes.  Lighthill refers to the low Reynolds number situation, which is appropriate for our purposes, as {\it RNW streaming} after the pioneers of its theory Rayleigh \cite{Rayleigh}, Nyborg \cite{Nyborg}, and Westervelt \cite{Westervelt}.

 Acoustic streaming is the result of forcing by a gradient in the {\it Reynolds stress} associated with high frequency (acoustic) waves in the fluid. The gradient results from attenuation of the acoustic waves.
  For our purposes, the basic equations of motion are the time-averaged Stokes equations,
 \begin{equation}
 0= \bar{F}_{j}-\partial \bar{p}/\partial x_{j}+\mu \nabla^{2}\bar{u}_{j},\hspace{.1in} \partial\bar{u}_{j}/\partial x_{j}=0
 \end{equation}
 where
 \begin{equation}
 \bar{F}_{j}=\partial (\rho \overline{u_{i}u_{j}})/\partial x_{i}.
 \end{equation}
 Here $\vu$ is the oscillatory velocity field and $p$ is the corresponding pressure.  The bar indicates the time average taken at a point over many cycles so that $\overline{\vu}$ is the streaming velocity we seek. The non-linear quantity  $\rho \overline{u_{i}u_{j}}$ is the Reynolds stress which represents the mean momentum flux associated with the acoustic wave. Its gradient is a force which is non-zero as long as some mechanism for sound attenuation is present. In the acoustic streaming models it is this force that is responsible for self propulsion.

 The attenuation necessary for streaming can occur in the body of the fluid or in a thin {\it Stokes boundary layer} surrounding a surface. Acoustic streaming due to attenuation in the body of the fluid can be observed in the laboratory when a quartz crystal is electrically excited to create high frequency vibrations. The ultrasonic beam from a face of the crystal, generated by the piezoelectric effect, can can create a turbulent jet in air with velocities in the tens of centimeters per second \cite{Lighthill}. This form of acoustic streaming  is commonly called a {\it quartz wind} (QW). In Lighthill's words,   ``not only can a jet generate sound,
but also sound can generate a jet''  \cite{Lighthill}. The QW effect is generally associated with high power sources of acoustic energy and  with very high frequency so that attenuation in the bulk of the fluid is significant.

  The second form of acoustic streaming occurs near boundaries.  Here the attenuation of the sound wave is a result of strong shear stresses within the  Stokes boundary layer that result from the no-slip boundary condition.   If $\vU$ is an oscillating velocity field in a fluid then the Stokes boundary layer is the thin layer surrounding the boundary where the vorticity is non-zero; $\vU$ is assumed to be irrotational outside this layer. The effective thickness of the Stokes boundary layer is $5(\nu /\omega )^{1/2}$ where $\nu$ is the kinematic viscosity and $\omega$ is the frequency of $\vU$ \cite{Lighthill}.  Boundary induced streaming can be generated either by a acoustic wave in the fluid or vibrations of the boundary itself; the streaming is a result of the relative motion. Vibrations of a solid leading to boundary induced acoustic streaming are  commonly called {\it surface acoustic waves} (SAW's) . Within the Stokes layer flow is turbulent and rotational, it can be regarded as a chaotic "atomosphere" surrounding the cell. Outside the Stokes layer flow is laminar and irrotational.

For a standing wave with membrane velocity $v(x)e^{i\omega t}$, Rayleigh's classical result states that the streaming velocity is $$\overline{u}_{s} = -\frac{3}{4\omega}v(x)v\hspace{.01in} ' \hspace{.01in} (x)$$
As an application we can use this to approximate the streaming velocity associated with the oscillatory motion of a yeast cell using the parameters found by Pelling \cite{Pelling}. If we take $v(x)=0.003 (1500\pi)\sin (2 \pi x)$ corresponding to a 1.5kHz vibration with a 3nm amplitude. We have (arbitrarily) taken the spatial wavelength to be one micron. The streaming velocity is approximately $-0.1\sin (2x)$$\mu$m/sec. We note that the streaming velocity is directed away from the antinodes and towards the nodes and is not propulsive. A progressive wave is necessary for acoustic streaming to be propulsive. We take up this situation in the next section.

 \section{Models}

 Here we propose two models for self-propulsion using acoustic streaming, the first based on the quartz wind effect and the second based on boundary induced streaming.

 \subsection{Quartz wind}  In this simplest model, the spicules  vibrate at a high frequency  buckling  the crystalline outer layer  in a manner similar to that of a piezoelectric door buzzer.  Attenuation of the acoustic beam in the bulk of the fluid generates a flow. Our original idea came from the Brazilian samba instrument known as a ``cuica'. (A cuica consists of a drum with a short bamboo reed penetrating its head. Vibrations in the reed are generated by rubbing it with a piece of cotton. The vibrations are transmitted to the drum head creating a loud noise.)  The spicules become active in a small region of the outer shell and the system works as an
 ``ultrasonic samba loudspeaker''.

 While the simplicity of this model is appealing, the acoustic power necessary for an organism to swim using this mechanism might be unrealistically high.
   Lighthill  has argued that the force ($F$) is related to the acoustic power ($P$) and the speed of sound in the fluid ($c)$ by  $F=P/c$ \cite{Lighthill2}. We can use this relationship to make a ``back of the envelope'' computation: Stoke's law $F= 6\pi \mu a v $ gives the force required to push a sphere of radius $a$ through a fluid with viscosity $\mu$ at speed $v$. The necessary acoustic power is then  $P  =  6\pi \mu a v c $. Synechococcus has a radius of about $10^{-4}$cm and swims in sea water with viscosity of $10^{-2}$g/cm\hspace{.005in} sec at about $2.5\times 10^{-3}$cm/sec. It would therefore require about
 $7\times 10^{-3} g\hspace{.01in} cm^{2}/sec^{3}$ or $7\times 10^{-10} $ watts of acoustic power to drive Synechococcus at observed speeds. By comparison, the power needed to push the cell is $6\pi \mu a v^{2} $ or approximately $1.18\times 10^{-17}$ watts.

Lighthill \cite{Lighthill3} defines an efficiency $\eta$ for a swimming mechanism as the ratio of the power required to push the cell to the power required by the mechanism:
\begin{equation}
\eta=6\pi \mu a v^{2}/P.
\end{equation}
 By this definition, the efficiency of the quartz wind mechanism is only about $1.7( 10^{-6})$\%.  A power output enhancement mechanism would need to be present to make this strategy biologically feasible\cite{Weaver}; see the discussion section below.

\subsection{Boundary induced streaming}
In this model, a high frequency traveling SAW is generated on the crystalline outer layer of the cell leading to boundary induced streaming. A standing SAW generates streaming towards the nodes of the wave but is not propulsive in an unbounded fluid.  The progressive wave induces a steady slip-velocity at the outer edge of the Stokes boundary layer. This flow creates the thrust necessary for propulsion. We find that the Lighthill efficiency for boundary induced streaming, which approaches 1\% and compares favorably with type C strategies (squirming) and flagellar propulsion. The amplitude of the SAW necessary to drive an organism the size of Synechococcus at observed speeds is on the order of $10^{-7}$cm, too small to be resolved using light microscopy.  The results of Pelling \cite{Pelling} and J\"{u}licher \cite{jul} indicate that the required frequencies are feasible and not unprecedented in nature.

 \subsubsection{Velocity, power, and efficiency for the SAW mechanism}  We estimate the velocity and efficiency for a spherical cell that swims using a traveling SAW. For simplicity, we restrict our attention to a tangential SAW though a normal SAW would also lead to self-propulsion. We compute the slip velocity at the edge of the Stokes boundary layer and the power (per unit area) for a swimming slab then use these results to estimate the swimming velocity and power output for a spherical organism. This approach was proposed in \cite{shapere} and provides a good estimate when the wavelength of the SAW is much smaller than the radius of the cell.

 Consider a slab of infinite extent with coordinates $(x,y)$ bounding an infinite region of water in the region $z\geq 0$. Suppose tangential progressive waves pass along the slab in the $x$-direction with velocity given by the real part of
\begin{equation}
U = A \omega \hspace{.01in} \mbox{exp}( i(n x-\omega t)) \,\,.
\end{equation}
 Longuet-Higgins \cite{Longuet1} derives the formula
\begin{equation}
U_{L} = \frac{5-3i}{4i\omega}(u_{1}^{(0)}-u_{1}^{\infty}) \frac{\partial}{\partial x}(u_{1}^{(0)}-u_{1}^{\infty})^{*})
\end{equation}
whose real part is the limiting streaming velocity at the edge of the Stokes boundary layer. (We have used * to indicate complex conjugate.)  Here $u_{1}^{(0)}\hspace{.01in} (=U)$ is the tangential velocity at the boundary and $u_{1}^{(\infty )}$ the solution to the linearized the Navier-Stokes equations evaluated just outside the Stokes boundary layer. For us $u_{1}^{(\infty )}=0$.  For the swimming slab we have
\begin{equation}
U_{L}=-\frac{5\pi}{2\lambda} \omega A^{2} \label{eq:UL}
\end{equation}
where $\lambda = 2\pi/n$ is the wavelength.

Now consider a spherical organism of radius $r$ that swims by passing high frequency traveling compression waves along its outer membrane. Let $(\theta,\phi)$ be spherical coordinates with $\phi$ the azimuthal coordinate measured from the front of the organism. Take the wave to be
\begin{equation}
\phi_{m}=\phi + \epsilon \sin (n\phi - \omega t)  \label{eq:wave}
\end{equation}
 where $\phi_{m}$ represents a material point on the outer membrane. The amplitude of the of the velocity is $A=\epsilon r \omega$ and the wavelength is $\lambda = 2 \pi r /n$. We assume that $\lambda <<r$ so that the local streaming velocity is well approximated by (\ref{eq:UL}).
Assuming the Stokes boundary layer to be of negligible width we use
\begin{equation}
\vU = \frac{5}{4}n \omega \epsilon^{2} {\bf i}_{\phi} \label{eq:slip}
\end{equation}
  as a slip velocity over the boundary of the organism.

 A convenient formula for the translational velocity associated with any boundary velocity field  was derived using the Lorentz reciprocal theorem in \cite{faire} and in \cite{Stone}, which, for a sphere is
\begin{eqnarray}
\vV = A(\vU )= -\frac{1}{4\pi r^{2}}\int \int_{S}\vU \hspace{.05in} d S \label{eq:connection}
\end{eqnarray}
where the integral is taken over the surface of the sphere.
Evaluating this with velocity (\ref{eq:slip}) we find that the spherical organism swims with velocity
\begin{equation}
\frac{5}{16}\pi n r \omega \epsilon^{2}
\end{equation}
along the axis of symmetry.  Note that the amplitude of the SAW is $r\epsilon$ and $\omega =2\pi f$ where $f$ is measured in Hertz. This velocity is 2.5 times that predicted by the squirming mechanism all parameters being equal \cite{Ehlersetal}.

\subsubsection{Efficiency comparisons}

To estimate the effort required to execute the compression waves we compute the power
\begin{equation}
{\cal P} = \int \int_{S} v_{i}\sigma_{ij}dS_{j} \label{eq:power}
\end{equation}
averaged over a swimming stroke. Again we assume $\lambda <<a$ and approximate the average power using the average power per unit area for a waving sheet. For a sheet in the $xy$-plane with a fluid of viscosity $\mu$ filling the region $z\geq 0$, the average power per unit area necessary to deform according to $x_{m}=x+A\sin (kx-\omega t)$ is
\begin{equation}
2\pi \mu \omega^{2}A^{2}/\lambda \label{eq:Psheet}
\end{equation}
where $\lambda =2\pi /k$, see \cite{Blake1} or \cite{Childress}. For the sphere deforming according to (\ref{eq:wave}) we have $A=r\epsilon$ and $\lambda=2\pi r /k$. Substituting these into (\ref{eq:Psheet}) and multiplying by the area we arrive at
\begin{equation}
{\cal P}=4\pi \mu r^{3} n \omega^{2}\epsilon^{2}.
\end{equation}
We note that this expression is in good agreement with the result obtained by evaluating (\ref{eq:power}) in spherical coordinates when $n\geq 10$.

The power output and efficiency for a cell of radius $10^{-4}$ using boundary induced acoustic streaming to swim at $2.5\times 10^{-3}$, the observed speed of Synechococcus,  is given in table 1.
We have (arbitrarily) chosen $n=30$. As we mentioned before, with efficiencies around 1\% this strategy is more efficient than the quartz wind strategy by many orders of magnitude, unless some power enhancement mechanism is present.

\begin{table*}
\caption{{\bf Efficiencies for the SAW mechanism}}
\centering
\begin{tabular}{llll}
frequency (Hz)& Amplitude (cm) & Power (Watts)  &$ \eta$ (\%) \cr
\\ \hline
500&$1.64\times 10^{-6}$&$1\times 10^{-15}$ &1.17\cr
1000&$1.16\times 10^{-6}$&$2\times 10^{-15}$   &  0.59 \cr
1500&$9.49\times 10^{-7}$&$3\times 10^{-15}$   &  0.39 \cr
5000&$5.20\times 10^{-7}$&$1\times 10^{-14}$& 0.12 \cr
\hline
\label{table:efficiencies}
\end{tabular}
\end{table*}

 \section{Discussion}
 The quartz wind strategy is less efficient by many orders of magnitude and is probably not biologically feasible unless some mechanism for power enhancement is present. On the other hand, all things being equal, propulsion by surface acoustic waves predicts a swimming velocity 2.5 times that predicted
by squirming. We believe it fair to say that, all things being equal, singers are faster than squirmers but slower than rowers.  For Synechococcus, the required frequency of the SAW is within the range observed in other biological systems (e.g. bacterial flagellar motors). The amplitude required for observed speeds is on the order of $10^{-7}$ cm, below the resolution limit of light microscopy. This leads to the key question. If acoustic streaming generated by surface acoustic waves is responsible for the locomotion of Synechococcus  how would one be able to "listen to their songs"?

We propose the following ideas, based on recent developments in micro-engineering, that could be objects of investigation.
We hope to stir interest in laboratory experimentation.

\subsection{Listening to the sound of cells: AS nanosensors}
One would like to be able to ``hear'' the sound generated by a 1 $\mu m$ moving
Synechococcus, via  nanosensors attached to the crystalline shell.
A clever way to do this is in order. Cantilever/nanowire  devices are already available
that can measure piezoelectric displacement transduction with frequency and amplitude ranges near the quantum regime \cite{Arlett}, \cite{Ekinci}.
Pelling, et. al  measured periodic oscillations with amplitudes of   3nm at frequencies of 0.8-1.6kHz on the of the outer membrane of Yeast cells using the cantilever of an atomic force microscope \cite{Pelling}. Living Yeast cell that measure about 5$\mu$m  in diameter were trapped in the micro-pores of a filter for the experiment. Metabolic oscillations were differentiated from Brownian motion by treating the cell with a metabolic inhibitor that does not change the mechanical properties of the cell wall. Treated cells did not display the oscillatory behavior observed in untreated cells. Yeast cells were chosen for the experiment due to their stiff cell wall; the spring constant of the cantilever needs to be comparable to the spring constant of the cells wall. Could this experiment be adapted to listen to a living Synechococcus?

 As for direct visualization of the local flow adjacent to a swimming Synechococcus, there are no technological limitations. By 2011 it is expected that
particles of 25 $nm$ be able to be manipulated/removed on chips, see  International Technology Roadmap for Semiconductors $http://www.itrs.net/$.

We believe it possible to map the flow pattern of the fluid adjacent to a swimming cell using a technology such as total internal reflection velocimetry (see the review by Guasto \cite{Guasto}). If so, this could be matched with the flow characteristic of acoustic streaming induced by surface acoustic waves. For a standing (but non-propulsive) SAW small particles would tend to collect at the nodes according to Rayleigh's theory. Detailed analysis of the fluid mechanics and careful experimentation would be required in the case of a progressive (propulsive) SAW. In our opinion this is a very interesting and challenging mathematical problem: to model the chaotic flow pattern inside the thin "atmosphere" (the Stokes layer) surrounding the cell. How should the organization and coordination of the molecular motors be in order to optimize the desired outcome, be it locomotion or nutrient uptake?

For the state of the art on AS sensors at the micro realm see the recent
review paper \cite{Lindner}.  Certainly the techniques of this paper could be applied to the raphe of a diatom skeleton to probe for
possible piezo-mechanical properties. Fluorescent beads would be focused by a standing SAW
\cite{Shi}. The ultimate challenge is to make measurements on a living cell.
Dynamical pressure measurements along the micrometer-sized channel  of the raphe of a living
diatom could be made using a micro-mamometer \cite{Abkarian}.   Could these ideas be used in the nano realm?
Piezoresponse force microscopy is used for imaging nanostructures \cite{Rodriguez} but it is not clear if it could be used to probe SAW emissions from a (dead or alive) diatom or a Synechococcus.

\subsection{Quartz wind  enhancement: uasers and submicrobubbles}
Quartz wind is a very simple mechanism, but  a very high metabolic rate
would be required and might not be biologically feasible.  One way  to remedy this drawback is to imagine a power enhancement mechanism similar to a laser.  {\it Uasers}  \cite{Weaver} are coupled ultrasonic  transducers producing stimulated emission via
positive feedback with an internal power mechanism. Power output scales with the square of the number of oscillators.  One could try to verify the signs of phase locked excitations.
 Figs. 3-8 of \cite{Hashimoto} invite looking for equivalent
biological structures in the cell membrane.

 Acoustic streaming acting on submicrobubbles (\cite{Attard}, \cite{Koishi}) attracted to the CS could produce streaming flows
   \cite{Marmottant1}, \cite{Marmottant}, \cite{Longuet},   \cite{Dijkink}. Power enhancement results from resonance.
  Another possibility is that spicules may  act as ``bubble poppers'', so that the cell takes full
  advantage of an external source of mechanical energy (\cite{Blake}, \cite{Katz}, \cite{Lugli}).

  Since the 1970's, power source systems for intracellular transport (molecular motors), locomotion systems for bacteria
(electro-chemical rotary motors), and protozoa (distributed dynein motors along the axonemes) have been identified.
 Ratchet vs. power stroke camps divided the scientific community for some specific systems.    Ratchet mechanisms are now
 taken seriously, as they have been reconciled
 with thermodynamics  \cite{Nikulov}. A locomotion model based on direct extraction of energy stored
in submicrobubbles may add fuel to the controversy, but we believe is a hypothesis worth exploring.
Micro heat engines, converting external heat sources to electrical energy, have been recently produced \cite{whalen}.

  \subsection{Hydrophylic/hydrophobic transitions} Micro-engineered surfaces coated with nanonails, when charged,  exhibit controlled  hydrophylic/hydrophobic transitions  \cite{nanonails}. This is suggestive, since the spicules project 0.15 $\mu$m   to the exterior of the crystalline shell.
 One can speculate that an hydrophylic-hydrophobic wave could entrain pumping motion, mediated perhaps by some ratchet type asymmetry or
 bubble manipulation.  Devices with chemically induced hydrophylic-hydrophobic microtracks have been recently constructed \cite{Renaudin1},\cite{Wixforth}.

 \subsection{Enhanced nutrient uptake}
We believe that AS is not just one more way of moving. It has been known since the fundamental work by Nyborg \cite{Nyborg1} that local mixing near the boundary is enhanced by AS.

Recent experimental literature seems to confirm
that AS enhances local mixing  \cite{Lutz},  \cite{Rathgeber}, \cite{Sritharan},  \cite{suria}.  Pelling has noted that the sound produced by yeast cells may be an indication of a pumping system that supplements passive diffusion. We would suggest that the sound itself may be the pumping mechanism.

 Synechococcus was an early lab-on-a-chip chemist.  Fundamental in the ecological chain, cyanobacteria were
the inventors of photosynthesis 2 billion years ago \cite{Liang}.  In the process of bubble collapse, several chemical reactions
occur \cite{suslik}, \cite{Suslik1}.  A curious coincidence is that chemical reactions involving nitrogenous compounds
are commonly produced in bubble cavitation \cite{Lohse}.  This may be of interest since cyanobacteria are indeed attracted to nitrogen.
In \cite{Magar} the average mass transfer available to a  spherical ``squirming swimmer''
(using tangential surface waves) is estimated. An important parameter here is the P\'eclet number, governing the ratio between advection
to diffusion. It would be interesting to compare this with estimates of mass transfer and mixing
coming from acoustic streaming locomotion processes. Perhaps some controlled laboratory experiment
could be devised using  chemo-attractants that would react near a Synechococcus.

\section{Epilog}
Nearly 50 years after Richard Feynman's lecture ``There's plenty of room at the bottom'' \cite{Feynman}, nanoengineering has advanced to the point where its developments could, in retribution,  benefit theoretical biology.
We have proposed  that acoustic streaming might be involved in the locomotion of the cyanobacterium Synechococcus.    To our knowledge the propagation of high frequency (acoustical) waves has never been proposed as a means of bacterial self-propulsion.
Actually a biological model for the ``cuica effect'' on a fluid was
 developed by Jackson and Nyborg \cite{Nyborg1958} already in 1958.   They were thinking of the reverse of the effect proposed here, namely, an external source (such as a  randomly vibrating microbubble)
 generating intracellular processes. In hindsight this could provide a recharging mechanism for the power systems in the cell. (We wish to provide, just as a historical correction, to the Physical Acoustics Timeline, 550 BC - present
(asa.aip.org/physical/patimeline.pdf). It is reported there that in 1960 ``Dyer and Nyborg describe studies in which a localized divergent sound source is brought into contact with the cell wall and suggest that intracellular motions are related to acoustic streaming and can be explained in terms of acoustic streaming theory.''  Actually it was 1958, one year before Feynman's lecture.)

Feynman would  certainly love to see a mechanical microprocessor \`a la
Babbage \cite{Roukes}  constructed with structures like that of the Synechococus CS-Sp complex. At this point it is perhaps worth ending with an often  quoted, but never over-quoted, excerpt from \cite{Feynman}:
\medskip 
{\small \begin{quote}
 ``The biological example of writing information on a small scale has inspired me to think of something that should be possible. Biology is not simply writing information; it is doing something about it. A biological system can be exceedingly small. Many of the cells are very tiny, but they are very active; they manufacture various substances; they walk around; they wiggle; and they do all kinds of marvelous things---all on a very small scale. Also, they store information...  The kind of writing that I was mentioning before, in which I had everything down as a distribution of metal, is permanent. Much more interesting to a computer is a way of writing, erasing, and writing something else.''
\end{quote}

}

\section*{Acknowledgments} We thank Howard Berg, Richard Montgomery, John Bush, and Sandra Azevedo for giving their encouragement and comments. KE acknowledges the support from mathematics and physics  faculty at St. Mary's College of Maryland where this research was conducted, and JK acknowledges the partial support from Nehama G.A. Our interest in the subject began in the mid 1990's
and we thank the Brazilian agencies CNPq, CAPES and FAPERJ for several grants allowing the scientific interchanges since then.

\end{document}